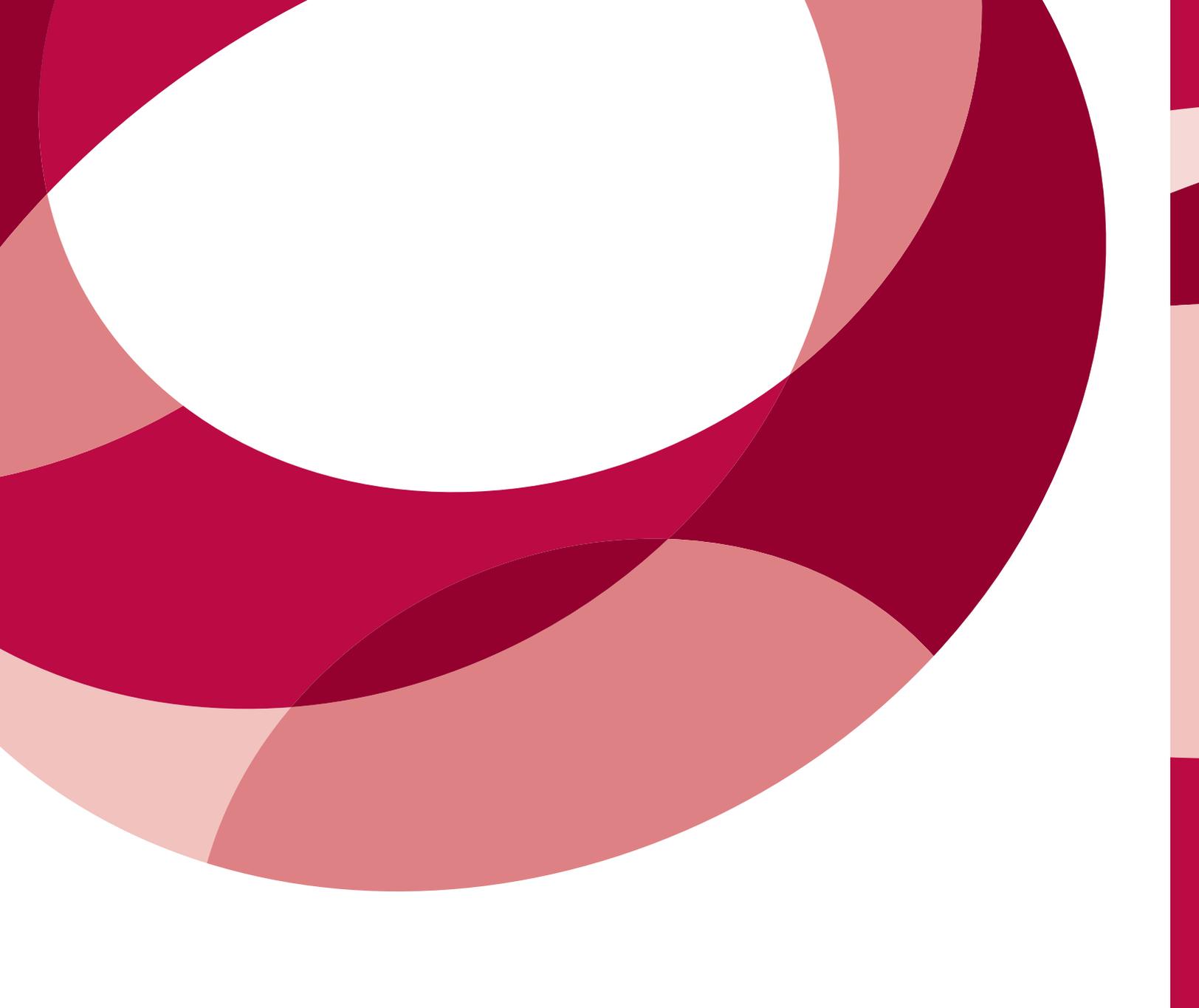

# Next Steps in Quantum Computing: Computer Science's Role

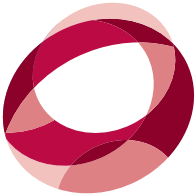

CCC
Computing Community Consortium
Catalyst

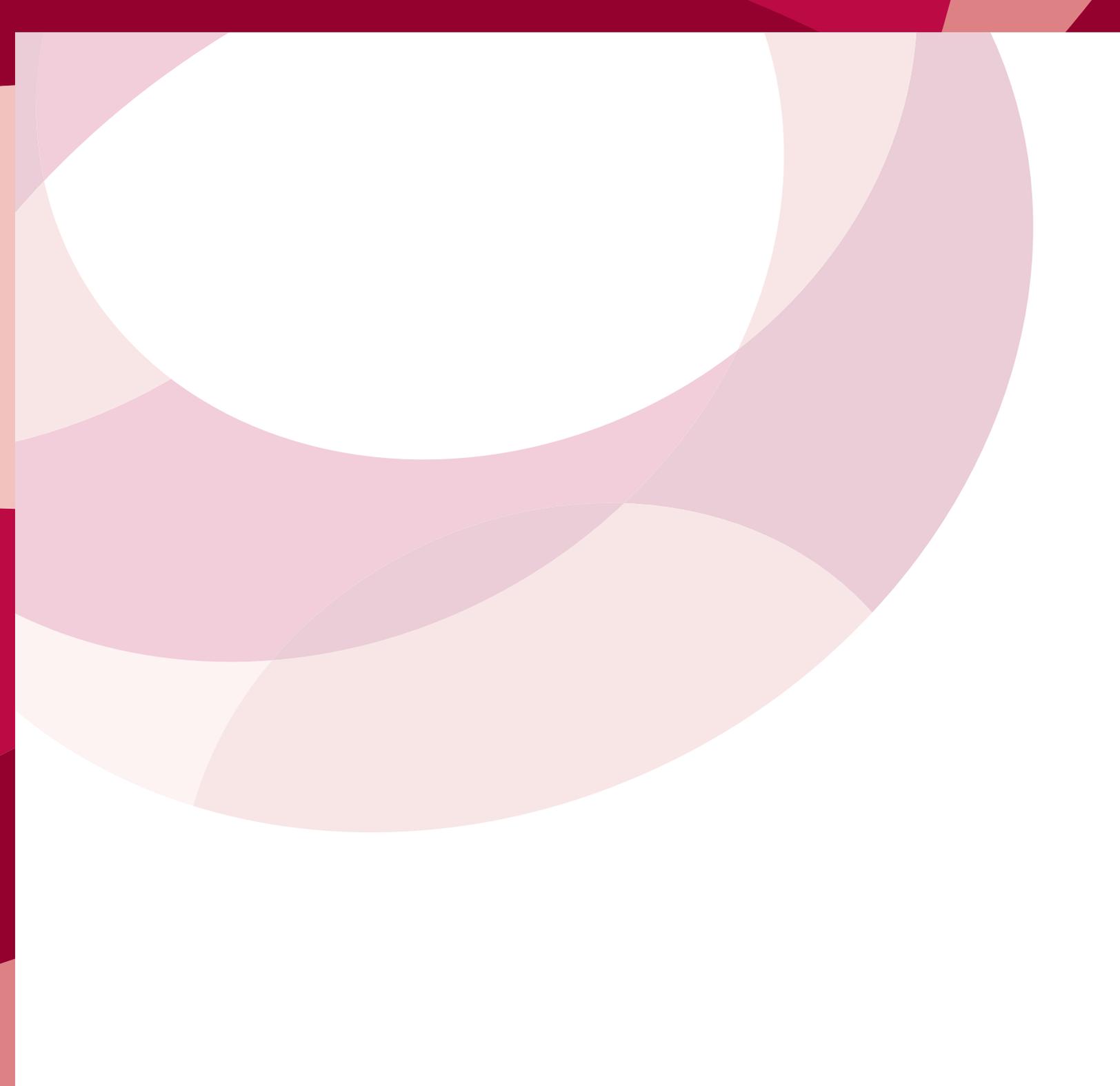

This material is based upon work supported by the National Science Foundation under Grant No. 1734706. Any opinions, findings, and conclusions or recommendations expressed in this material are those of the authors and do not necessarily reflect the views of the National Science Foundation.

# Next Steps in Quantum Computing: Computer Science's Role

Margaret Martonosi and Martin Roetteler, with contributions from numerous workshop attendees and other contributors as listed in Appendix A.

November 2018

Sponsored by the Computing Community Consortium (CCC)

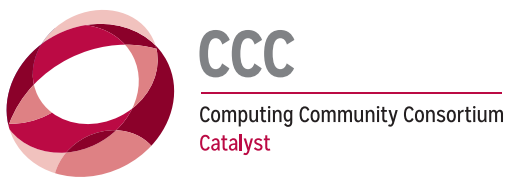

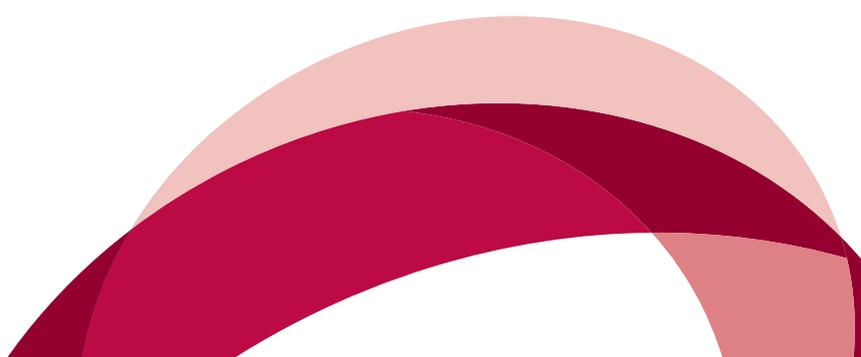

# 1. Introduction

The computing ecosystem has always had deep impacts on society and technology and profoundly changed our lives in myriads of ways. Despite decades of impressive Moore's Law performance scaling and other growth in the computing ecosystem there are nonetheless still important potential applications of computing that remain out of reach of current or foreseeable conventional computer systems. Specifically, there are computational applications whose complexity scales super-linearly, even exponentially, with the size of their input data such that the computation time or memory requirements for these problems become *intractably large to solve for useful data input sizes*. Such problems can have memory requirements that exceed what can be built on the most powerful supercomputers, and/or runtimes on the order of tens of years or more.

**Why Quantum Computing?** Quantum computing (QC) is viewed by many as a possible future option for tackling these high-complexity or seemingly-intractable problems by complementing classical computing with a fundamentally different compute paradigm. Classically-intractable problems include chemistry and molecular dynamics simulations to support the design of better ways to understand and design chemical reactions, ranging from nitrogen fixation[1] as the basis for fertilizer production, to the design of pharmaceuticals[2,3]. Materials science problems that can be tackled by QCs include finding compounds for better solar cells, more efficient batteries, and new kinds of power lines that can transmit energy losslessly[4]. Finally, Shor's algorithm[5], which harnesses QC approaches to efficiently factor large numbers, raises the possibility of making vulnerable the current data encryption systems that rely on the intractability of this calculation; the existence of a QC sufficiently large and sufficiently reliable to run Shor's on full-length keys could make current cryptosystems vulnerable to attack and eavesdropping.

**What is Quantum Computing?** QC uses quantum mechanical properties to express and manipulate information as quantum bits or qubits. Through specific properties from quantum physics, a quantum computer can operate on an exponentially large computational space at a cost that scales only polynomially with the required resources. Algorithms that can be appropriately implemented on a quantum computer can offer large potential speedups — sometimes even exponential speedups — over the best current classical approaches.

QC therefore has the potential for speedups that are large enough to make previously-intractable problems tractable. For instance, on a classical computer, it would take quadrillions of years to find the ground state energy of a large molecular complex to high precision or to crack the encryption that secures internet traffic and bitcoin wallets. On a quantum computer, depending on the clock-speed of the device, these problems can potentially be solved in a few minutes or even seconds.

**The Inflection Point: Why now?**
The intellectual roots of QC go back decades to pioneers such as Richard Feynman who considered the fundamental difficulty of simulating quantum systems and "turned the problem around" by proposing to use quantum mechanics itself as a basis for implementing a new kind of computer capable of solving such problems . Although the basic theoretical underpinning of QC has been around for some time, it took until the past 5 years to bring the field to an inflection point: now small and intermediate-scale machines are being built in various labs, in academia and industry[7 8]. Preskill has coined[9] the phrase Noisy Intermediate-Scale Quantum (NISQ) to refer to the class of machines we are building currently and for the foreseeable future, with 20-1000 qubits and insufficient resources to perform error correction[10]. Increasingly, substantial research and development investments at a global scale seek to bring large NISQ and beyond quantum computers to fruition, and to develop novel quantum applications to run on them.

---

[1] https://arxiv.org/abs/1605.03590
[2] https://arxiv.org/abs/1808.10402
[3] https://arxiv.org/abs/1706.05413
[4] https://www.nature.com/articles/s41467-017-01362-1 or http://spie.org/newsroom/6386-quantum-techniques-to-enhance-solar-cell-efficiency?SSO=1
[5] https://arxiv.org/abs/quant-ph/9508027v2
[6] https://link.springer.com/article/10.1007/BF01886518
[7] https://www.nature.com/articles/nature08812
[8] http://www.pnas.org/content/114/13/3305
[9] https://arxiv.org/abs/1801.00862

**The Research Need:** There is a huge gap between the problems for which a quantum computer might be useful (such as chemistry problems[11], material science problems, etc.) and what we can currently build, program, and run. As Figure 1 conceptually illustrates, many well-known QC algorithms have qubit resource requirements that far exceed the current scale at which QCs can be built.

The goal of the QC research community is to close the gap such that *useful* algorithms can be run in *practical* amounts of time on *reliable real-world QC hardware*. Although the current pace of development is high, the projected time to close this algorithms-to-machines gap is still often viewed as ten years or more in the future. Current efforts in QC are focused on accelerating research and development in order to close the gap sooner. In particular, the goal of this Computing Community Consortium (CCC) workshop was to articulate the central role that the computer science (CS) research communities plays in closing this gap. CS researchers bring invaluable expertise in the design of programming languages, in techniques for systems building, scalability and verification, and in architectural approaches that can bring practical QC from the future to the present.

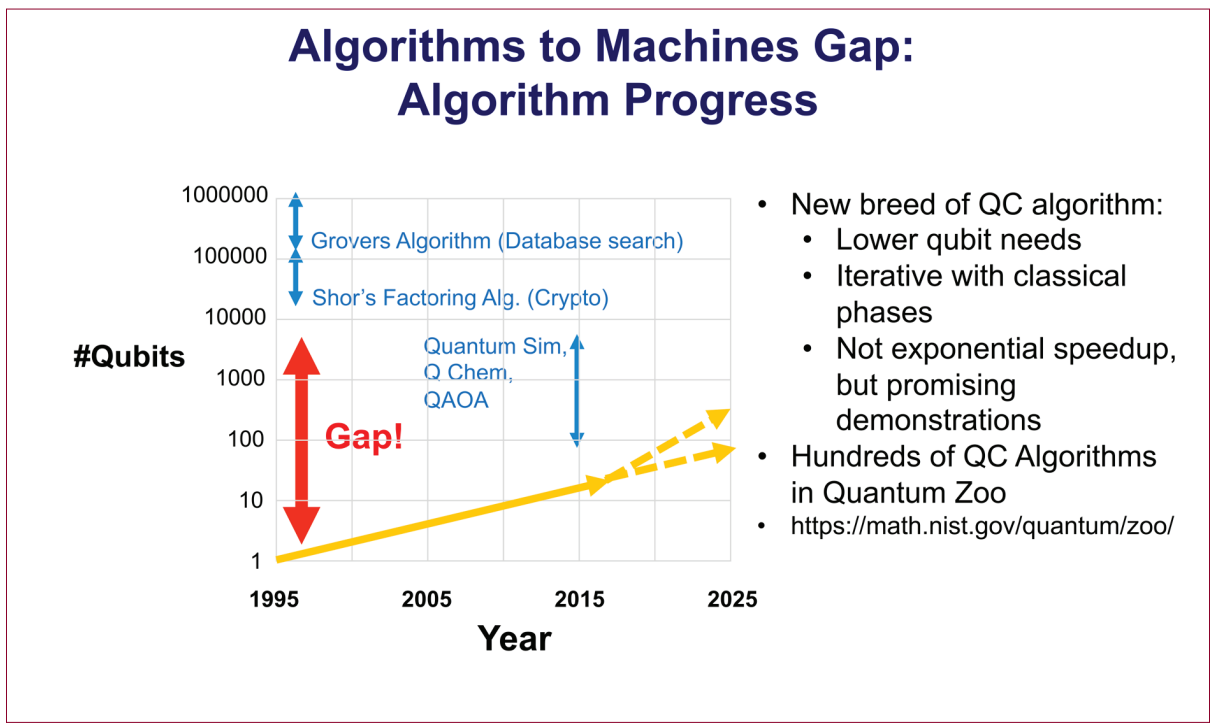

*Figure 1: The Algorithms-to-Machines gap illustrates how well-known QC algorithms (such as Shor's and Grover's) have resource requirements that far exceed the qubit counts (shown in yellow) of systems we are able to build.*

---

[10] https://quantum-journal.org/papers/q-2018-08-06-79/
[11] https://journals.aps.org/pra/abstract/10.1103/PhysRevA.90.022305

This report introduces issues and recommendations across a range of technical and non-technical topic areas:

◗ **Quantum Computing Algorithms:** What are QC systems good at? Can we catalyze the development of end-to-end QC algorithms that offer compelling speedups over classical?

◗ **QC Technologies: How will we build QC systems?** In some ways, QC today represents classical computing circa 1950. Building sufficiently large and reliable QC systems will require substantive advances in the underlying QC technologies themselves, as well as the software layers to support them.

◗ **Programming environments and toolchains for QC:** One of the key opportunities for QC advances lies in identifying methods to facilitate the ability of domain experts and algorithmicists to express QC applications and their constraints using high-level abstractions, and then to compile and map these onto QC hardware or simulators. While QC toolchains already exist, improving the support of high-level abstractions and strong verification and optimizations can dramatically lessen the hurdles faced by QC algorithms developers.

◗ **QC Architectures:** Finally, techniques drawing from computer architecture research could help develop QC systems that transition QC from the current approach of undifferentiated qubit arrays, towards more differentiated organizational structures. Techniques that can help exploit locality and exploit particular technology attributes and characteristics can offer substantial leverage.

**Broader Cross-Cutting Issues:** In addition to the technical imperatives above, Computer Science's role in advancing QC research also hinges on broader issues. These include ensuring that the broader CS research community sees a role for itself. On the following pages we provide a summary of our recommendations for resources that should be created in order for QC CS to advance as a field.

**Research Recommendations:**

### CCC Next Steps in Quantum Computing Workshop: Overview of Research Needs

**Classical and Quantum Algorithms:**

First and foremost, there is an overarching need for new Quantum Computing algorithms that can make use of the limited qubit counts and precisions available in the foreseeable future. Without a "killer app" or at least a useful app runnable in the first ten years, progress may stall.

Given the potential scientific and commercial promise of Quantum Chemistry algorithms (e.g. agriculture, pharmaceuticals, and basic research), participants felt that the field will benefit from further Quantum Chemistry algorithm innovations and hardware capabilities sufficient for more sophisticated models, such as simulating the properties of excited states and dynamics as well as ground states.

Although QC implementations sufficient to execute Shor's algorithm on practical key sizes are many years away, research is needed in advance of that on "Post-Quantum" public-key cryptographic systems that can resist quantum attack and maintain security.

**Programming, Compilation, and Systems Layers:**

The workshop agreed that there is a general need for research regarding how best to implement and optimize programming, mapping, and resource management for QC systems through the functionality in between algorithms and devices.

For near-term NISQ machines, we will need to create and refine languages and compilation techniques that give programmers the expressive power needed to articulate the needs of QC algorithms relative to tight resource constraints on current implementations.

Longer term, the use of abstractions to enhance productivity will be needed, once quantum resources are more plentiful. (For example, effective QEC techniques may allow future QC programmers to treat QC instructions and qubits as fully reliable and accurate.) We must establish the sorts of modularity and layering commonly needed for scalable systems. These include libraries for commonly-used functions, as well as APIs and instruction sets to aid development and optimization.

Quantum debugging is a fundamental challenge, both because measuring a qubit collapses its state, and also because underlying implementations have operator precision errors and yield probabilistic results. Therefore, research is needed regarding how we can track the errors that accumulate through a quantum program, how we can verify that the computation operates within a tolerated error, and how we can facilitate the process of QC software debugging?

Computer architectures for QC systems will need to be researched and experimented with. What sort of microarchitectures and functional units work best now, and how will they scale to larger systems? Likewise, work is needed to explore qubit communication and transportation issues, such as teleportation units, particularly as qubit counts scale beyond what local communication can support for larger QCs.

Real-world quantum systems will be hybrids of classical and quantum units. Research is needed on how to program and map efficiently to "both sides" of such machines. Opinions vary on the degree of architectural sophistication warranted on each side, as well as on issues of communication between them.

As with other systems and architecture research, the development of hardware and software techniques must be paralleled by the simultaneous development of metrics to define performance and reliability, and evaluation infrastructure to estimate, simulate, or measure them.

**Quantum Implementations:**
The "winning technology" is currently far from clear. The field needs to continue to innovate on fabrics for quantum technologies based on different physical device approaches. In particular, implementation advances will hinge not just on device physics, but also on close collaboration between interdisciplinary teams of computer scientists and physicists to advance QC hardware organizations and approaches overall.

Efficiency in QC Linear systems and machine learning algorithms hinges on finding an efficient way for the quantum hardware to access large amounts of classical input data, which is currently a fundamental bottleneck for potential speedups in this application class.

Being able to produce high-dimensional superpositions containing either the relevant input data or some nonlinear function of it will facilitate performance improvements in tasks such as clustering, PCA, and other data analysis tools for machine learning and optimization.

The participants identified the opportunities for error reductions and precision improvements in existing and near-term QC systems, including through applications of machine learning to machine data and error characterizations.

Given underlying hardware error rates, QC systems will use quantum error correction (QEC) to achieve lower overhead and lower thresholds when resources permit their implementation. Research is needed to identify the most promising QEC implementations, particularly ones that can support the state of qubits over long periods of time and long sequences of operations.

Current NISQ systems do not have sufficient resources to implement QEC. In time however, QC implementations will need the capacity to generate, consume, and recycle a high volume of clean auxiliary qubits which will be a fundamental support for all current QEC approaches in large QC machines.

As with classical computing, the memory system plays a crucial role. Participants noted the need for research on QC memory system design, such as the basic challenge of storing the instructions and state required for a large number of qubits at ultra-low temperatures (kelvin/sub-kelvin)?

**Conferences and Community Resources:**
The participants felt there was a need to develop conferences and communities to help people working on different parts of the QC "stack" to share approaches and to interact with a wide range of applications and devices specialist.

QC research will benefit from involving people and ideas from many other fields. Particularly mentioned were topics like probabilistic programming and the approximate/unreliable computing field, for instance recent work on program logics for union bound and verifying quantitative reliability.

In addition to conference interactions, the community will benefit from producing shared APIs and standard interface layers that allow toolchains and approaches from different academic or industry entities to interoperate with each other. Likewise, where languages, compilers, and software systems can be open-sourced, this will better support full-stack advances from applications and programs down to device specifics.

## 2. Workshop Methods

The CCC Next Steps in Quantum Computing workshop was held on May 22-23, 2018, in Washington D.C. It brought together researchers from quantum computing with experts in other fields of computer science, such as computer architecture, electronic design automation, compiler construction, and classical programming languages to discuss how computer scientists can contribute to the nascent field of quantum computing.

The workshop was focused around four main topic areas: algorithms, technologies, toolchains/programming, and architecture. Discussion in each area was kicked off by two speakers who presented on its current state within the context of quantum computing. The speakers for each area were:

◗ Algorithms: Andrew Childs (University of Maryland) and Xiaodi Wu (University of Maryland)
◗ Technologies: Andrew Houck (Princeton University) and Jungsang Kim (Duke University)
◗ Toolchains/Programming: Bettina Heim (Microsoft Research) and Ali Javadi-Abhari (IBM Research)
◗ Architecture: Igor Markov (University of Michigan) and Fred Chong (University of Chicago)

Each of the presentations were followed by discussion with the entire group and then participants were divided into four breakout groups for more in-depth discussion. Breakout groups then presented the conclusions from their conversations for the entire group of participants for additional comments. On the afternoon of day two, participants used the materials from the breakout groups and group discussions to begin drafting this report. A full agenda of the workshop can be found online at https://cra.org/ccc/events/quantum-computing/.

The full list of workshop participants can be found in the appendix.

## 3. Technology Trends and Projections

Projecting trends in QC is challenging, and there is no consensus on whether a "Moore's Law for Quantum" will emerge. Moore's Law refers to the long-term trend in classical computing by which advances in semiconductor fabrication would lead to regular (every 18-24 month) doublings in the number of transistors that could be cost-effectively fit on a single chip[12]. Moore's Law has been sustained for decades due to a "virtuous cycle": these scalings in transistor counts, performance, and functionality provide the revenue that subsidizes the research/technology advances needed for the next technology generation. Unlike Moore's Law for classical semiconductors, there is no positive prognosis for a virtuous cycle of QC technology scaling; beginning such a cycle would require a wider range of useful QC applications at low-enough resource requirements to implement and see speedups from today. Nonetheless, the ability to forecast and reason about different scaling options remains a useful topic for guiding R&D progress, so portions of the workshop discussed methods for quantifying advances in QC, and even for "roadmapping" their progress.

When QC is discussed both in the news and in many broad-interest technical publications, the primary figure of merit portrayed is often qubit count. Indeed the number of physical qubits successfully built and tested is an important indicator of the computational power of the machine. On the other hand, focusing single-mindedly on qubit counts to the exclusion of other important metrics can be problematic. For example, qubits do not hold their state indefinitely, so the coherence interval (i.e., how long their state is held) becomes an important figure of metric. In addition, qubit movements, operations, and readouts all have some element of imprecision, so different qubit error rates are also important.

Figure 2 (next page) illustrates a possible 2D technology scaling space. Physical qubit counts lie on the y-axis, while the physical error probability lies on the x-axis. Technology developers make design choices that lie on different points

---

[12] https://ieeexplore.ieee.org/abstract/document/591665/

## Noisy, Intermediate Scale Quantum (NISQ) Computers

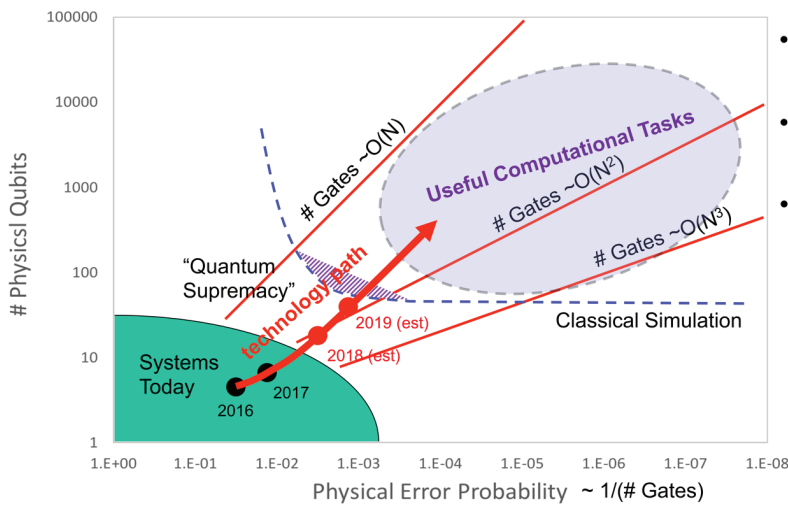

- First, reach a fault-tolerant qubit
- Then scale up in numbers
- Interesting computational tasks beyond classical simulation limit

*Figure 2: The possible scaling trajectories of the number of physical qubits vs physical error probability.*

in this space. For example, some might choose to hold qubit counts constant while working to improve physical error rates of the machine. Conversely, others might hold error rates roughly constant while scaling qubit counts. While qubits get the headlines, it is likely that some combination of more qubits and more reliable qubits/gates will be a more useful scaling trajectory.

A key focal point for the QC community pertains to so-called quantum advantage (also referred to as "quantum supremacy"[13,14,15] or "quantum computational supremacy"[16] by some authors). This is the point where QC first exceeds the performance of classical computing for some particular problem. The specifics of what will constitute an initial quantum advantage demonstration are still being discussed, because ongoing advances in classical computing simulations make quantum advantage a moving target[17]. Figure 2 depicts a blue line that represents a possible crossover trend. Namely, it will require both

"enough" qubits and also that each qubit or gate operation is sufficiently reliable. If quantum advantage is the initial and most pressing goal for QC, then teams will select scaling strategies that they believe will get them "across the blue line" first.

In addition to these relatively-short-term scaling discussions, there are also longer-term issues to be considered. For example, there was a strong sense at the workshop that end-to-end performance metrics should be the primary ruler by which QC performance is gauged. Namely, some QC algorithms offer high speedups under the assumption that QC state has been previously initialized in memory. Efficient Quantum RAMs and methods for initializing them have not yet been identified, meaning that some such algorithms will not offer speedups once the time (sometimes exponential) for state preparation is accounted for. Other figures of merit for QC include aspects of the resource requirements or complexity of the applications and algorithms.

---

[13] https://arxiv.org/abs/1203.5813
[14] https://www.nature.com/articles/s41567-018-0124-x
[15] https://arxiv.org/abs/1807.10749
[16] https://arxiv.org/abs/1809.07442
[17] https://arxiv.org/abs/1807.10749  [17] https://arxiv.org/abs/1807.10749

*CS Research priorities and recommendations:* Several key roles for CS researchers emerge from these technology and scaling trends:

◗ *Resource estimation tools to project resource requirements for different algorithms and mapping approaches.*

◗ *Intelligent mapping and scheduling approaches that make the best use of scarce QC hardware resources.*

◗ *Effective and resource-efficient error correcting code strategies to overcome low-reliability qubits or operations.*

## 4. Algorithms

Ultimately, the promise of quantum computers is the ability to run a qualitatively different form of algorithm which can solve some problems far more quickly than any possible classical computer.[18] Determining which problems would benefit from a quantum computer is still an active area of research, and one of crucial importance for CS researchers. The known quantum algorithms have been compelling enough to stimulate enormous interest and investment in the field, but continued progress will require both improving these algorithms and coming up with new ones[19,20].

In classical computing, our empirical understanding of algorithms is far ahead of our theoretical understanding. For example, one of the most useful classical algorithms is MCMC, which has been in use since the 1940s, but has only recently been proved to be correct in nontrivial cases, and even these proofs establish performance bounds that are usually far worse than what is empirically observed. The study of quantum algorithms has so far been dominated by theory work since we have not had access to large quantum computers that can be used to test algorithmic ideas. However, we expect that as QC hardware matures, the field of quantum algorithms will become increasingly empirical in its focus, and eventually will rely on the mix of theory, heuristic, and observation that we see in classical computing.

### 4.1 Cryptanalysis

Shor's algorithm[21] was an early milestone in QC algorithms as it demonstrated the possibility for QCs to offer exponential speedup on a problem of practical interest — namely, factoring large numbers. Many current cryptosystems, such as RSA, rely on the fact that factoring large (e.g. 2048 bit) numbers into their component primes is intractably hard. If a QC were built with sufficient resources and reliability to run Shor's algorithm on 2048-bit keys, it could potentially decrypt information previously considered secure[22]. As already envisioned by Shor in his 1994 paper, a similar attack can be mounted[23] against the so-called discrete log problem. This will break the authentication behind most of the currently used cryptocurrencies, including Bitcoin, Ethereum, and others, as well as other blockchain technologies. While these potential attacks require high-quality quantum bits, it is known that a moderate number of them are sufficient to break RSA (around 4000 error-corrected qubits for 2048-bit keys) and Bitcoin encryption (around 2300 error-corrected qubits for 256-bit keys).

As such, there is a major effort underway to find "post-quantum" public-key cryptosystems that will resist quantum attack. Some of the leading candidates are based on lattices, for which no efficient quantum algorithms are known yet. The main way that we gain confidence in the security of a cryptosystem is by attacking it. Thus there is an urgent need to study possible quantum algorithms for lattice-based as well as code-based cryptosystems in order to find out whether these too will turn out to be vulnerable. Unlike other quantum algorithms, here it is important to know whether an algorithm (say for breaking lattice cryptosystems) will be possible long before the algorithms are implemented. This is because updating software takes time, especially for embedded systems, and because an adversary could save encrypted messages today and decode them later when large quantum computers become available.

---

[18] https://www.nature.com/scientificamerican/journal/v298/n3/full/scientificamerican0308-62.html
[19] https://arxiv.org/abs/0808.0369
[20] https://www.nature.com/articles/npjqi201523
[21] https://dl.acm.org/citation.cfm?id=264406
[22] https://cacm.acm.org/magazines/2013/10/168172-a-blueprint-for-building-a-quantum-computer/fulltext
[23] https://arxiv.org/abs/1706.06752

Because Shor's algorithm would require many orders of magnitude more resources than are currently available, it is not yet runnable on numbers intractable to factor classically[24,25,26]. Thus, a key focus of QC algorithms work today is on identifying algorithms of practical interest that can make practical use of the size and reliability of QCs available now during the NISQ era. The subsequent sections discuss these possibilities in more detail.

## 4.2 Quantum simulation

Feynman's original vision of quantum computing hinged on its value for simulating complex quantum mechanical systems, and this remains an area of active interest. For decades, conventional computer simulations have expanded our understanding of quantum mechanical systems, but the complexity of these simulations has forced them to employ approximations that ultimately limit the amount of useful information we can extract. The basic difficulty is the same fact that makes quantum computers effective: describing a quantum system requires a number of parameters that grows exponentially with the size of the quantum systems. Quantum computers are ideally suited to simulate quantum mechanical systems in a variety of disciplines, including quantum chemistry, materials science, nuclear physics, and condensed matter physics.

The Boston Consulting Group has estimated that improved quantum simulation could have a market value of tens of billions of dollars to pharmaceutical companies alone[27]. Quantum simulation (including chemistry, lattice QCD, material science, etc.) currently accounts for a large fraction of supercomputer time, and we expect that quantum computers would not merely be able to accomplish these simulations more cheaply but also greatly expand the range of what is possible with them.

Several quantum simulation algorithms have already been proposed and tested on quantum computers[28,29,30]. These initial algorithms have been designed for systems requiring minimal resources[31,32]. One promising current line of research is hybrid quantum-classical approaches. These approaches off-load certain computations onto classical computers, e.g. Hamiltonian integrals can be pre-computed on a classical computer and then loaded into the quantum computer algorithm as parameters. Conversely, a quantum computer could be used to speedup critical parts in simulations, e.g., providing information about two-particle density matrices. Ground-state properties are typically obtained using variational methods[33,34]. These are iterative methods in which one chooses an initial wave function depending on one or more parameters, and then determine parameter values that attempt to minimize the expected energy values. The resulting wave function is an upper bound on ground state energy. Iteration (e.g. via gradient descent) can continue to improve the estimate.

In the future, we expect there to be a strong need for new algorithms as the number of qubits and available number of gate operations increase, because we will no longer be constrained to minimize resources. Quantum computers are expected to be able to simulate properties of excited states and dynamics as well as ground states. Most classical *ab initio* codes (i.e., those relying on basic natural laws without additional assumptions or special models) are limited to simulating static properties of ground states. There is also a need for new transformations mapping particle systems obeying either fermionic and bosonic statistics onto registers of distinguishable quantum bits that might be constrained by particular hardware connectivities[35].

---

[24] https://dl.acm.org/citation.cfm?id=2638690
[25] https://dl.acm.org/citation.cfm?id=3179560
[26] https://dl.acm.org/citation.cfm?id=2011525
[27] https://www.bcg.com/en-sea/publications/2018/coming-quantum-leap-computing.aspx
[28] https://www.ncbi.nlm.nih.gov/pubmed/16151006
[29] https://www.nature.com/articles/nature23879
[30] https://journals.aps.org/prx/abstract/10.1103/PhysRevX.8.011021
[31] https://www.nature.com/articles/nchem.483
[32] https://arxiv.org/abs/1001.3855
[33] https://www.nature.com/articles/ncomms5213
[34] https://www.nature.com/articles/nature23879
[35] https://arxiv.org/abs/1710.07629

Beyond physics simulations themselves, there are also opportunities in related topic areas including protein modelling, molecular dynamics, weather prediction, fluid mechanics, drug design, and computational optics. By employing QCs for the classically-intractable portions of commercially-relevant problems in drug design or other fields, QCs of sufficient scale and reliability have the potential for significant commercial relevance.

### 4.3 Machine Learning and Optimization

Much less is known about the utility of quantum computers for machine learning, but the importance of the application makes this a compelling area of study. If we can produce high-dimensional superpositions containing either the relevant input data or some nonlinear function of it, then we can quickly perform clustering, PCA, and other data analysis tasks. However, this initial state preparation is still a hurdle. Obtaining a useful overall speedup will require preparing a state of $2^n$ dimensions in much less than $2^n$ time, preferably in poly(n) time. We currently only know how to do that in some special cases[36]. It would be of great utility to expand the range of cases where this is possible.

Variational and adiabatic algorithms for optimization and classification can run on near-term quantum computers and yet are out of reach of classical simulators[37]. Although these approaches show promising advantages in quantum chemistry and simulation[38], they have not yet provably outperformed the best known classical algorithms. Empirical evidence from running them on near-term quantum computers will improve our understanding for longer-term and more scalable approaches.

### 4.4 Quantum Error Correction (QEC)

Current NISQ systems are too resource constrained to support error correction, but the field still looks ahead to a future (10+ years away) where error correction schemes may be employed to support the state of qubits over long periods of time and long sequences of operations[39]. As such, to some involved in QEC research, QEC itself is the primary workload that will be running on QCs of the future[40,41,42]. Future research is required to develop QEC approaches that are effective and resource efficient, so that they can be employed sooner (ie at lower qubit counts) in the technology timeline.

### 4.5 Quantum Advantage

The Quantum Advantage milestone assumes specific quantum computation benchmarks[43] that would defy simulation on classical computers. Google has recently proposed such benchmarks for a certain sampling problem, spurring serious improvements in simulation algorithms. As a side effect, researchers found a number of loopholes in the benchmarks that make simulation much easier. Known loopholes have been closed when Google published revised benchmarks. In general, we expect there will be a period of cat-and-mouse as loopholes emerge and are closed again in quantum advantage benchmarks. As one example, sequences of diagonal gates should be avoided because they enable efficient tensor-network contraction methods. Computational-basis measurements applied after diagonal gates can also be exploited. In some cases, these and other loopholes have properties that can be checked by verification techniques.

### 4.6 CS Research Opportunities and Timeline for Quantum Algorithms

While hundreds of QC algorithms exist[44], there remains a fundamental need for applications and algorithms that offer useful speedups on the NISQ machines available now or soon.

**Promising in the near-term (roughly 10-20 years):** In this timeframe, NISQ machines with 1000 qubits and 4 9s (99.99%) of operator precision are expected to exist. There is evidence that even rudimentary quantum computers will be useful as quantum coprocessors in hybrid classical-quantum architectures. Applications include variational algorithms for quantum chemistry[45] and materials science, as well as other special-purpose simulators. In some cases, rigorous error bounds on simulation results can be given, for instance via the Quantum Approximate Optimization Algorithm (QAOA)[46].

---

[36] https://arxiv.org/abs/1408.3106 and https://arxiv.org/abs/1611.09347

[37] https://www.nature.com/articles/nature17658

[38] https://www.ibm.com/blogs/research/2017/09/quantum-molecule/

[39] https://arxiv.org/abs/0904.2557 [40] https://arxiv.org/abs/1809.01302

[41] https://dl.acm.org/citation.cfm?id=3123949

[42] https://dl.acm.org/citation.cfm?id=3123940

[43] https://arxiv.org/abs/1807.10749

[44] https://math.nist.gov/quantum/zoo/

**Special-purpose simulators:** These involve devices that are either not universal for quantum computing, or which are much more effective at simulation their "native" Hamiltonian than in applying a universal gate set. The main examples come from trapped ions, neutral atoms or molecules. The appeal here is such devices can be easier and cheaper to build than universal QCs, while there are also disadvantages such as systematic errors and lack of flexibility.

**Improve and make better use of existing algorithms:**
There is a lot of room for improvement in currently known quantum algorithms. For example, the linear systems algorithm[47,48] or machine learning would be vastly improved if there was a more efficient way for the quantum hardware to access classical data. Quantum search and amplitude amplification are general techniques that can be used to provide a polynomial speed-up, given that the classical algorithm can be efficiently reformulated in the quantum realm.

Variational algorithms, such as the ground state estimation previously discussed, are promising for the NISQ era but currently lack theoretical bounds on performance. Research on these bounds would allow prediction of the number of qubits and fidelities required to get useful output from these algorithms[49].

Running concrete examples on the real hardware that does exist will lead to efficient techniques that can make the difference between a practical algorithm or an unrunnable one. Doing so will require developing effective and near-optimal techniques mapping and scheduling to a particular gate set, connectivity, and control constraints. This co-design will accelerate the development of the overall quantum computing *system*.

The same theoretical algorithm may have several different physical operations that implement it, for example, different quantum circuits can achieve the same unitary transformation. Some of these will be more sensitive to noise and control errors than others. So far most error-analysis has been done at the gate level. Noise-analysis at a higher level will lead to more noise-resistant algorithms, which is essential in the NISQ era.

---

[45] https://www.nature.com/articles/ncomms5213
[46] https://arxiv.org/abs/1602.07674
[47] https://journals.aps.org/prl/abstract/10.1103/PhysRevLett.103.150502
[48] https://journals.aps.org/pra/abstract/10.1103/PhysRevA.89.022313
[49] https://dl.acm.org/citation.cfm?id=2685188.2685189

# 5. Devices

The landscape for QC devices is fast-changing, making it difficult to place a clear bet on a long-term winner. Table 1 shows a summary of technology options thus far. Given this report's focus on the CS role, we present the devices landscape for context, but do not elaborate deeply on the many important research needs embodied at this portion of the implementation "stack".

## 5.1 Current Status

In order to compete with classical computing, QC systems need both sufficient qubits and sufficient fidelity. The current state of the art, represented by superconducting circuits and trapped ions[52], are systems with approximately 10-50 qubits and approximately 1% error per gate ("2 9's" or 99% precision). Within 5-10 years, brute force efforts to scale these systems are likely to yield 100's of qubits with improved error rates. Such systems will be able to demonstrate quantum advantage—that is, a definitive performance improvement over a classical computer on a particular task. However, to solve challenging real-world problems, significantly more qubits and lower error rates will be required. Progress towards this goal may be achieved on several fronts, as previously illustrated in Figure 2:

1) Improve errors in existing systems through characterization and application of machine learning.

2) Improve error correcting codes to achieve lower overhead and lower thresholds.

3) Develop new fabrics for quantum technologies based on different physical systems. These may be radically different approaches, or hybrids of currently-studied platforms.

## 5.2 Devices Challenges and Opportunities

In NISQ systems, gate errors play a significant role and the performance of quantum computers will benefit significantly from cross-layer optimization. Physical fabrics and architectures will be very tightly tailored to the needs of specific implementations, with little to no resources expended for software-friendly abstractions. Once scalable, fault-tolerant logical qubits are developed, some systems may have sufficient reliability and resources to employ more abstraction can be deployed.

Advances in QC hardware will require close collaboration between interdisciplinary teams of computer scientists and physicists. For example, language, compiler and systems researchers are needed to allow expressive and optimizable toolflows down to hardware, accounting for resource constraints and non-idealities. Computer architects are needed to explore scalable and modular design approaches. For certain topics like machine learning or molecular dynamics simulations, we must explore problem-specific hardware in different physical platforms. Overall, these cross-cutting research efforts will be required to exploit the full suite of optimization opportunities for the system performance.

| Technology | Best Argument For | Best Argument Against | Companies Involved |
|---|---|---|---|
| Majorana | Fundamentally protected from errors | Hard to engineer | Microsoft |
| Solid-state spins (P:Si, NV centers, etc.) | Small footprint | Heterogeneous, hard to scale | Turing, CQC2T |
| Quantum dots | Small footprint, scalable fabrication | Connectivity | HRL, Intel |
| Neutral atoms | Homogeneous, long-range gates | Lack of demonstrated good 2-qubit gates | Atom Computing, Inc. |
| Linear optics[50] | Scalable fabrication | Lack of key components (single photon sources) | PsiCorp, Xanadu |
| Superconductors | Demonstrated programmability, lithographically definable | Large footprint, 10 mK | Google, IBM, Rigetti, Intel, QCI |
| Ions[51] | Demonstrated programmability, long coherence, homogeneous, | Microsecond gate speeds, lasers | IonQ, Honeywell |

*Table 1*

---

[50] https://www.nature.com/articles/nphys2253

[51] https://www.nature.com/articles/nature18648

[52] https://www.nature.com/articles/nature18648

# 6. Architecture

## 6.1 Overview

As in the classical domain, the ultimate role of QC computer architecture is to organize and abstract physical resources in ways that improve modularity, aid in design correctness, simplify programming, and enhance flexibility for machine evolution between generations. At this early period, some architectural exploration is warranted to develop computer organizations that work well under current NISQ constraints[53]. Given the fast-moving changes in device technologies, longer-term architectural choices may not yet be obvious, but exploration of QC architectural concepts should begin now nonetheless, with an eye towards future, large-scale quantum computers.

One fundamental abstraction in classical (non-QC) computer architectures is the instruction set architecture (ISA), which describes a set of software-visible operations that all implementations of that ISA must implement, albeit in different ways. Given current resource scarcity (i.e., 100s of qubits or less), it is important to optimize across the computing stack. In this era, abstraction is eschewed; the tight optimization of specific resources makes it undesirable to have the equivalent of an ISA. This environment also favors application-specific architectures (e.g., those addressing quantum chemistry problems, Hamiltonian dynamics simulations problems, or classical-quantum hybrid algorithms that leverage variational quantum eigensolvers or quantum approximate optimization ansatz) over more generalized architectures. Furthermore, organizations that exploit locality and parallelism may be important for enhancing performance of QC programs, and even for enhancing the likelihood that they progress to completion without error accumulation rendering their answer uselessly inaccurate. Also, reduced resource circuits could be obtained based on optimizes libraries of quantum circuits[54].

In the future when resources are more plentiful, abstractions become feasible and can enhance productivity. An obvious abstraction is that effective QEC techniques may allow future QC programmers to treat QC instructions and qubits as fully reliable and accurate; that is, with sufficient resources, error correction is abstracted away by the ISA and hidden in the microarchitecture. Similarly, alternative gate implementations can be handled in the translation / optimizing compilation process rather than hand optimized. Traditional CS topics such as programming languages, back-end code generation, dynamic runtimes and potentially virtualization will be extremely helpful in supporting this transition.

Classical circuits and computation elements have multiple roles in quantum computer architectures. In particular, architects should consider the architecture for control of quantum circuits (e.g. pulse generators) and should also consider the interface between classical and quantum computation.There is a need for multiple levels of program specifications, from very hardware centric[55] to those closer to a universal ISA[56,57] for quantum computation. A future push toward characterizing quantum algorithms running on particular hardware must consider full end-to-end (Classical Input->Compute->Classical output) aspects and performance (quality of the computational output and/or time taken to complete a given computational task). In the near/intermediate term, such characterization would involve a simulator, potentially of a high-performance kind, at all levels.

---

[53] https://dl.acm.org/citation.cfm?id=2011926
[54] https://ieeexplore.ieee.org/document/6983057/
[55] https://arxiv.org/pdf/1809.03452.pdf
[56] https://arxiv.org/abs/1707.03429
[57] https://arxiv.org/abs/1608.03355

Once quantum computing machines become large enough to utilize error correction, the support for error correction becomes a substantial aspect of the architecture. In particular, error correction requires a high volume of clean auxiliary qubits to be generated, consumed, and recycled. Further, control circuitry for performing error correction operations (measurements, correction operations, etc) ideally exist close to the quantum data path, since they represent a dominant aspect of the datapath operations[58]. This suggests that portions of the error correction logic should be implemented in classical logic compatible with cold domains and close to the quantum gates (to avoid frequent communication between classical and quantum environments)[59, 60, 61].

Parallelism at various levels will be an important consideration for near-term and future systems. Proper architectural support for parallel implementations, broadly defined, of quantum gates may be pivotal for harnessing the power of NISQ devices[62]. Different gate control primitives, e.g., local vs. global, may be exploited to enhance quantum computational power. When a quantum computer is used as an accelerator, concurrency among parts of a hybrid system must be understood and managed. On a larger scale, distributed multi-quantum systems will introduce new challenges.

In general, QC architectures face a three-dimensional tradeoff space of parallelism vs. coherence time vs. error optimization. The choices are technology specific (e.g., superconducting approaches suffer from memory error, thus parallelism is favored, whereas ion trap approaches favor lower gate counts). What type of layout makes sense? What type of layout is resilient to faulty links and/or variability in error rates of the links?

## 6.2 Communication

For larger QCs, communication will be a multi-scale problem. As qubit counts scale beyond what local communication can support, computer architectures will need to include resources specifically supporting communication, such as teleportation units[63]. Qubits themselves will be divided into modules in hierarchical interconnection domains, with varying technology and bandwidth. Such organizations will introduce a need to consider intra-module communication (often accomplished via swaps or ballistic movement) and inter-module (often accomplished via so-called quantum teleportation with distribution networks of Einstein–Podolsky–Rosen (EPR) pairs). Architecting for modularity and for efficient communications becomes an important aspect of the overall design; it impacts performance, reliability, scalability, and cost. As one example, logic for generating and distilling EPR pairs at opposite ends of long-distance communications will become important to optimize.

Likewise, there will be design decisions pertaining to the physical placement of resources. For example, some of the classical control circuitry that manages and sequences quantum operations will often benefit from being designed to operate in the cold (single-digit Kelvin) environment preferred by most current QC technologies. Architectural and design decisions that assess these locality and hot-cold tradeoffs will make interesting research directions for years to come.

## 6.3 Beyond the Qubits: Architectural Support for the Control and Memory modules

What type of microarchitecture is required to scale to a large number of qubits? How do we design a memory system to keep the instructions and state required for a large number of qubits and is still useful at ultra-low temperatures (Kelvin/sub-Kelvin)? This will likely mean that we have to manage data carefully to respect the varying power budgets (and thermal budgets) available at different domains. The control circuitry typically requires a million X hardware for providing microwave pulses and measurements. Is there a way to make this scale? On a related front, do we need hard-real time systems of some sort to feed instructions to the quantum datapath, or are simple state machines sufficient?

---

[58] https://dl.acm.org/citation.cfm?id=3123949

[59] https://dl.acm.org/citation.cfm?id=3123940

[60] https://ieeexplore.ieee.org/document/7927104/

[61] https://dl.acm.org/citation.cfm?id=3123952

[62] https://dl.acm.org/citation.cfm?id=2694357

[63] https://arxiv.org/pdf/1809.01302.pdf, https://people.cs.uchicago.edu/~ftchong/papers/Javadi-Abhari.pdf

## 6.4 QC Architectural Design: A Quantitative Approach

Across many of the issues described above, there is a strong need to develop evaluation infrastructure and metrics to define performance and reliability. As foundational metrics and tools become available, their presence will help in reaching out to wider architecture community. Models and simulations are essential for components, subsystems, and systems, allowing for experimentation and evaluation that anticipates hardware capabilities. System resources, such as storage, auxiliary qubits, and T-factories, must be considered in the overall architecture. System-level metrics are needed, that can be validated against actual implementations. Such cost metrics will serve as targets for optimizations.

In general, many computer architecture techniques revolve around exploiting non-uniformity at different scales, and quantitative tradeoffs of this sort will fuel interesting QC architecture research. For example, we know that qubits (and their connections) have different error rates. Can we design architectures that can exploit that variability, rather than being limited by the worst case qubit or link? In addition to tools, architects need characterization data about error rates to drive such solutions; it will be important to ensure that such datasets become available to a broad set of researchers.

## 7. Programming Models and Toolchains

One of the central roles for computer scientists in QC will be in the research and development of programming models and toolchains[64]. As illustrated in FIgure 3, QC today to some degree resembles 1950 in classical computing, with algorithm and device development outpacing the maturation of systems and programming functionality "in the middle". While classical computing adopted layering strategies in the 50's, 60's and 70's that to some degree have carried forward to today, QC may well take a different approach due to severe resource constraints in this context.

A first wave of programming and toolchain development occurred over the past ten years[65, 66, 67, 68] but much of that work focused attention on high-resource applications and algorithms far beyond NISQ capabilities. Current work must focus on techniques that give programmers the expressive power needed to articulate the needs of QC algorithms relative to tight resource constraints on current implementations.

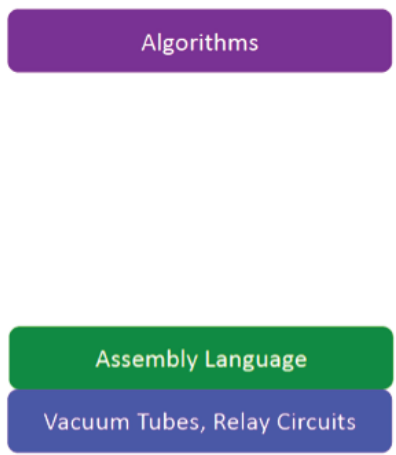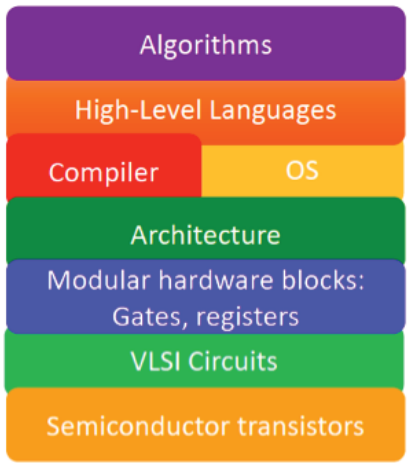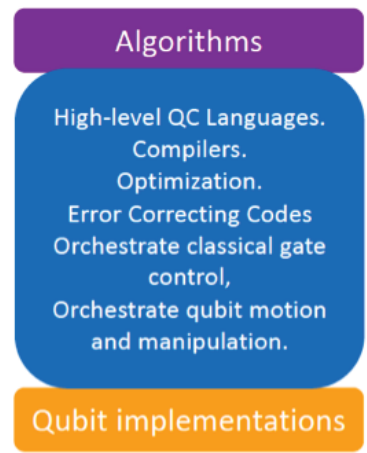

*Figure 3: Illustration of the layers involved in classical computing circa 1950s, current classical computing and quantum toolflows.*

---

[64] https://arxiv.org/pdf/1809.03452.pdf
[65] https://arxiv.org/abs/1402.4467
[66] https://dl.acm.org/citation.cfm?id=2462177
[67] https://dl.acm.org/citation.cfm?id=3183901
[68] https://dl.acm.org/citation.cfm?id=2597939

To some degree, the current state of QC programming and system layers resembles that of classical computing circa 1950. In particular, while research attention has been placed on algorithms and devices, more attention will need to be devoted to the functionality in between, in order to determine how best to implement and optimize programming, mapping, and resource management for QC systems. This is analogous to the software stack present in today's classical computers, though may well take a different form in QC due to the exceptional resource constraints at play.

In time, when resources permit, additional work will be needed on the sorts of modularity and layering commonly needed for scalable systems[69, 70]. For example, libraries for commonly-used functions will aid development and optimization[71]. Modules for key aspects of QEC are likewise important. Ultimately, one key question is to what degree it will ever be possible to program QCs without a deep or physical understanding of qubits, analogous to today's classical computing ecosystem where few programmers are well-versed in how transistors work.

In the short term, QC systems require heavy full-stack sharing of information and data from applications and programs down to device specifics. As such, languages and software systems will benefit from methods to aggregate and share such information. Likewise, and in more human terms, it will also be important for conferences and communities to be developed to help the people working on different parts of the toolchain to share approaches and to interact with a wide range of applications and devices specialist. In the longer term, standard interface layers might allow toolchains and approaches from different academic or industry entities to interoperate with each other.

In addition to programming and compilation, there are also runtime systems and operating systems to consider. Given the heavy control complexity of QC systems, important research will revolve around good methods for qubit calibration and for adapting to specific system characteristics. Further work is also needed regarding the dynamic coordination between quantum and classical parts of the program execution.

## 8. Verification

From the complexity of quantum algorithms to the unreliability of quantum gates, quantum computing is rife with the certainty of error. This calls for verifying every stage of quantum computation, from the programs used to generate quantum circuits to the hardware design. A *verified quantum computing stack* (akin to Princeton's Verified Software Toolchain[72]) would ensure that each level of the quantum computing process corresponds to a specification, improving reliability in the final system and enabling us to diagnose errors as they appear.

### 8.1 High-level: Quantum Programs

The highest level of this verification stack relates to quantum programs. The properties of quantum systems we may wish to verify range from lightweight to comprehensive. On the lightweight side, traditional programming language techniques like type systems and abstract interpretation allow us to verify properties like no-cloning (qubits cannot be copied), or the separability of two qubits (whether or not qubits are entangled). More heavyweight systems leverage the fact that quantum programs are well understood (they have a precise semantics in terms of superoperators) to verify arbitrary properties of circuits or fully characterize their behavior. These tools, like the QWIRE circuit language[73] in the Coq proof assistan[74], trade increased verification power for time and effort by sophisticated quantum programmers. Program logics like Quantum Hoare Logic[75] (implemented in Isabelle) and others can simplify the task of full program verification. Tools are also being developed to verify the specific class of quantum programs that

---

implement security protocols, like Unruh's Quantum Relational Hoare Logic[76] (modelled on the EasyCrypt cryptographic tool[77]). For the subset of quantum programs that consists of reversible problems, there are further tools available, for instance ReVerC[78] which is a compiler from F# to Toffoli networks that has been for which verified in the proof assistant F*.

## 8.2 Mid-level: Verifying accumulated errors

A major question for program verification is how we can track the errors that accumulate through a quantum program and verify that the computation operates within a tolerated error. Such errors may occur from the inherent noise in the computer, or the probabilistic nature of the algorithm. Tools such as type systems which can track and quantify the amount of error in a computation and report that information to the user could greatly help with the design of new algorithms, and to increase the reliability of programs which use the limited quantum processing time. This may be a good opportunity to bring in people and ideas from the approximate/unreliable computing field, for instance recent work on program logics for union bound[79] and verifying quantitative reliability (e.g., the Rely programming language[80]).

## 8.3 Hardware-Level

Adapting a quantum circuit to a given hardware architecture entails a variety of powerful optimizations related qubit placement, gate scheduling and pulse-sequence generation. These optimizations can reduce susceptibility to device errors and even route around known-bad qubits. The presence of QEC circuits adds another rich layer to possible optimizations. However, each category of optimizations brings possible bugs and adds to the demand for verification tools. In conventional Electronic Design Automation, the simplest verification category is formal circuit-equivalence checking. Relevant methods are based on Boolean SATisfiability, SMT solvers and Binary Decision Diagrams. These techniques remain formally applicable to Boolean reversible circuits, although verifying hardware-optimized modular exponentiation (for Shor's algorithm) remains computationally challenging. Quantum circuits with non-Boolean gates make SAT, SMT, and BDD techniques less applicable, even though there have been attempts[81]. There is also significant difference between exact equivalence and approximate equivalence - the latter does not imply an equivalence relation and thus breaks various computational methods. Approximate equivalences can also accumulate errors, so are more relevant for end-to-end evaluation of circuits and systems. Exact equivalence is relevant to local circuit transformations that are safe to compose in large numbers.

## 9. Conclusions

Overall, QC is poised at a deeply fascinating inflection point. Large industry and government investments are pushing for breakthroughs in qubit counts and fidelities, with quantum advantage being a much-sought-after milestone. To reach long-term practicality, however, will require considerable innovation after quantum advantage has been reached. Practical QC algorithms that can make use of intermediate-scale hardware will likely be needed in order to motivate ongoing investment of time and resources into QC developments. Without a "killer app" or at least a useful app runnable in the first ten years, progress may stall. In addition, the workshop agreed that there is a general need for research regarding how best to implement and optimize programming, mapping, and resource management for QC systems through the functionality in between algorithms and devices. Attention to systems design and scalability issues will be important as QC systems grow beyond small qubit counts and require modular large-scale designs. For near-term NISQ machines, we will need to create and refine languages and compilation techniques that give programmers the expressive power needed to articulate the needs of QC algorithms relative to tight resource constraints on current implementations. Longer term, the use of abstractions to enhance productivity (e.g. effective QEC techniques may allow future QC programmers to treat QC instructions and qubits as fully reliable and accurate) once quantum resources are more plentiful. We must establish the sorts of modularity and layering commonly needed for scalable systems (e.g. libraries for commonly-used functions, as well as APIs and instruction sets will aid development and optimization). Furthermore, real-world quantum systems will be hybrids of classical and quantum units, and research is needed on how to program and map efficiently to "both sides" of such machines. Opinions vary on the degree of architectural sophistication warranted on each side, as well as on issues of communication between them.

---

[76] https://arxiv.org/abs/1802.03188

[77] https://www.easycrypt.info/trac/

[78] https://github.com/msr-quarc/ReVerC or https://arxiv.org/abs/1603.01635

[79] https://arxiv.org/abs/1602.05681

[80] https://mcarbin.github.io/rely/

[81] https://arxiv.org/abs/1704.08397 or https://ieeexplore.ieee.org/document/1269084/

That being said, the "winning technology" is currently far from clear. The field needs to continue to innovate on fabrics for quantum technologies based on different physical device approaches. In particular, implementation advances will hinge not just on device physics, but also on close collaboration between interdisciplinary teams of computer scientists and physicists to advance QC hardware organizations and approaches overall - for instance collaboration with probabilistic programming and the approximate/unreliable computing field. The community will also benefit from producing shared APIs and standard interface layers that allow toolchains and approaches from different academic or industry entities to interoperate with each other. Likewise, where languages, compilers, and software systems can be open-sourced, this will better support full-stack advances from applications and programs down to device specifics. Across all the envisioned needs for QC's success, the engagement of the CS research community and the education of a QC-aware CS workforce will be important factors in achieving the needed research goals.

## 10. Appendix

**Workshop Attendees:**

| First Name | Last Name | Affiliation |
|---|---|---|
| Matthew | Amy | University of Waterloo |
| Kenneth | Brown | Duke University |
| Greg | Byrd | North Carolina State University |
| Jonathan | Carter | Lawrence Berkeley National Laboratory |
| Vipin | Chaudhary | National Science Foundation |
| Andrew | Childs | University of Maryland |
| Fred | Chong | University of Chicago |
| Almadena | Chtchelkanova | National Science Foundation |
| Dave | Clader | Johns Hopkins University Applied Physics Laboratory |
| Tom | Conte | Georgia Institute of Technology |
| Sandra | Corbett | Computing Research Association |
| Nathalie | de Leon | Princeton University |
| Khari | Douglas | Computing Community Consortium |
| Ann | Drobnis | Computing Community Consortium |
| Monisha | Ghosh | National Science Foundation |
| Markus | Grassl | Max Planck Institute for the Science of Light |
| Emily | Grumbling | National Academy of Sciences |
| Daniel | Gunlycke | U.S. Naval Research Laboratory |

| Aram | Harrow | Massachusetts Institute of Technology |
| Peter | Harsha | Computing Research Association |
| Mark | Heiligman | IARPA |
| Bettina | Heim | Microsoft Research |
| Mark | Hill | University of Wisconsin-Madison |
| Andrew | Houck | Princeton University |
| Meghan | Houghton | National Science Foundation |
| Travis | Humble | Oak Ridge National Laboratory |
| Sabrina | Jacob | Computing Research Association |
| Ali | Javadi-Abhari | IBM Research |
| Sonika | Johri | Intel Labs |
| Jamil | Kawa | Synopsys, Inc. |
| Jungsang | Kim | Duke University |
| Vadym | Kliuchnikov | Microsoft Research |
| John | Kubiatowicz | University of California at Berkeley |
| Brad | Lackey | University of Maryland, College Park |
| Yi-Kai | Liu | NIST and University of Maryland |
| Igor | Markov | University of Michigan |
| Margaret | Martonosi | Princeton University |
| Dmitri | Maslov | National Science Foundation |
| Anne | Matsuura | Intel Labs |
| Mimi | McClure | National Science Foundation |
| Michael | Mislove | Tulane University |
| Yunseong | Nam | IonQ |
| Massoud | Pedram | University of Southern California |
| Irene | Qualters | National Science Foundation |
| Moinuddin | Qureshi | Georgia Institute of Technology |
| Robert | Rand | University of Pennsylvania |
| Martin | Roetteler | Microsoft Research |
| Neil | Ross | Dalhousie University |
| Amr | Sabry | Indiana University |
| Peter | Selinger | Dalhousie University |
| Peter | Shor | Massachusetts Institute of Technology |
| Burcin | Tamer | Computing Research Association |
| Jake | Taylor | OSTP |
| Himanshu | Thapliyal | University of Kentucky |
| Jeff | Thompson | Princeton University |
| Heather | Wright | Computing Research Association |
| Helen | Wright | Computing Community Consortium |
| Xiaodi | Wu | University of Maryland, College Park |
| Jon | Yard | University of Waterloo |
| Kathy | Yelick | Lawrence Berkeley National Laboratory |
| William | Zeng | Rigetti Computing |

**Other contributors to the workshop report:** Prakash Murali (Princeton University), Teague Tomesh (Princeton University), Yipeng Huang (Princeton University).

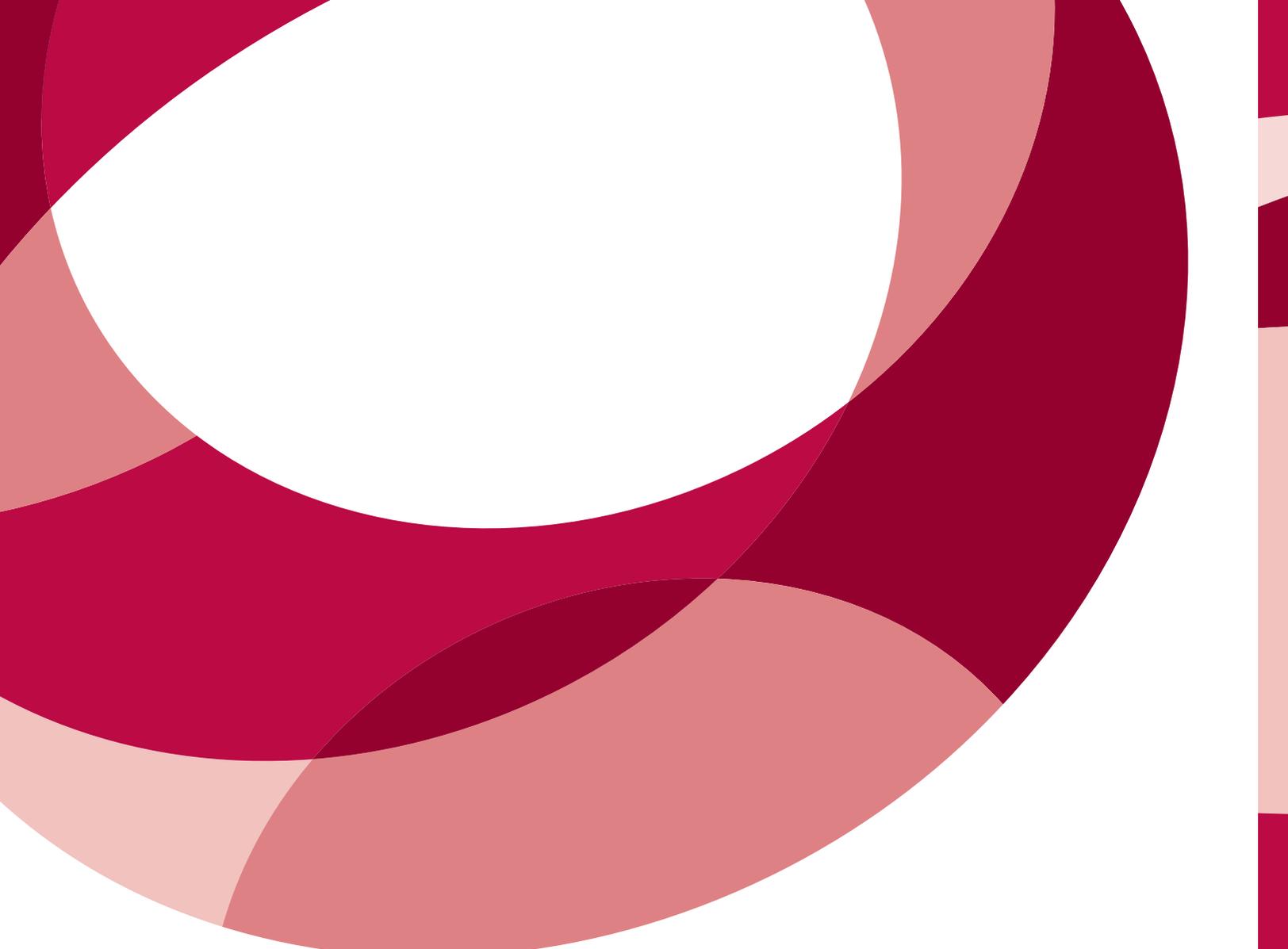
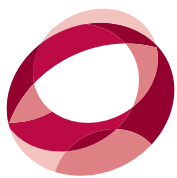

**CCC**
Computing Community Consortium
Catalyst

1828 L Street, NW, Suite 800
Washington, DC 20036
P: 202 234 2111 F: 202 667 1066
www.cra.org cccinfo@cra.org